\documentclass[12pt,a4paper]{article}
\usepackage{amsmath,epsfig}
\begin{document}
 
\setlength{\unitlength}{.8mm}
\newcommand{\be}{\begin{equation}}
\newcommand{\bea}{\begin{eqnarray}}

\newcommand{\Eps}{\Epsilon}
\newcommand{\gM}{\mathcal{M}}
\newcommand{\dD}{\mathcal{D}}
\newcommand{\gG}{\mathcal{G}}
\newcommand{\pa}{\partial}
\newcommand{\eps}{\epsilon}
\newcommand{\La}{\Lambda}
\newcommand{\De}{\Delta}
\newcommand{\nonu}{\nonumber}
\newcommand{\beq}{\begin{eqnarray}}
\newcommand{\eeq}{\end{eqnarray}}
\newcommand{\ka}{\kappa}
\newcommand{\an}{\ensuremath{\alpha_0}}
\newcommand{\bn}{\ensuremath{\beta_0}}
\newcommand{\dn}{\ensuremath{\delta_0}}
\newcommand{\al}{\alpha}
\newcommand{\bm}{\begin{multline}}
\newcommand{\fm}{\end{multline}}
\newcommand{\de}{\delta}

\begin{titlepage}
\begin{flushright}
MPI-PhT/03-19\\
April 2003
\end{flushright}
\vspace{1cm}

\begin{center}
{\LARGE \bf Test of asymptotic freedom and scaling hypothesis 
in the 2d O(3) sigma model}
\vspace{2cm}

{\large J. Balog$^1$ and P. Weisz$^2$}\\[7mm]

{\small\sl $^1$Research Institute for Particle and Nuclear Physics}\\
{\small\sl H-1525 Budapest 114 Pf. 49, Hungary}
\\[2mm]
{\small\sl $^2$Max-Planck-Institut f\"{u}r Physik}\\
{\small\sl D-80805 Munich, Germany}
\vspace{1cm}

{\bf Abstract}
\end{center}
\vspace{-5mm}

\begin{quote}
The 7--particle form factors of the fundamental spin field of the O(3) 
nonlinear $\sigma$--model are constructed. We calculate the corresponding
contribution to the spin--spin correlation function, and compare 
with predictions from the spectral density scaling hypothesis. 
The resulting approximation to the spin--spin correlation function 
agrees well with that computed in renormalized (asymptotically free) 
perturbation theory in the expected energy range. 
Further we observe simple lower and upper bounds for the sum of the 
absolute square of the form factors which may be of use for 
analytic estimates.

\end{quote}
\vfill
\end{titlepage}

\section{Introduction}
$1+1$ dimensional integrable models are theoretical laboratories where
one can study selected problems in field theory. These include 
the question whether the continuum limit of the lattice regularized
version of a field theory coincides with the theory constructed by
continuum techniques. One can study here
structural questions related to asymptotic freedom, topological excitations
etc., and, in the case that the theories agree, the nature  
of lattice artifacts.

The form factor bootstrap method \cite{FFB,FFaxioms} 
of constructive field theory has been used to construct many integrable 
models. The method consists of three main steps. 
The starting point is the 
exact scattering matrix that is calculated (sometimes conjectured) by 
the bootstrap method \cite{ZZ}. Next the form factors (matrix elements
of local operators between scattering states) are calculated by
solving the form factor (FF) axioms \cite{FFaxioms}. Finally correlation
functions are constructed by inserting a complete set of intermediate
states between the local operators. 

The program has so far only been carried out completely for the Ising model
~\cite{Ising}. The S--matrix is known for many integrable models and
if there are no internal degrees of freedom also the many--particle
form factors are easily constructed as a multiple product of the
basic two--particle form factor (which is known again for most models).
On the other hand, for models with internal degrees of freedom the
solution of the FF axioms, matrix functional difference equations in
this case, is not known in general. Nevertheless, the form factors
of some models belonging to this class, the Sine--Gordon model,
the chiral Gross--Neveu model, and the O(3) and O(4) nonlinear
$\sigma$--models, are available \cite{FFaxioms,Smirnov}. 
Finally (with the exception
of the Ising model) the $r$--particle contributions to the
correlation functions can only be computed numerically and therefore
this sum has to be truncated after the first few terms.

Among models with internal (isospin) degrees of freedom, it is
the O(3) model, where one can go furthest. This is because 
in this special case the many--particle
form factors (after removing a simple factor) are polynomials in the
particle rapidities which can be determined recursively \cite{BH}.
Fortunately it is also a model which is of particular interest since it 
has many properties akin to Yang--Mills theory in 4 dimensions in that it 
exhibits asymptotic freedom and has instanton solutions. 

The question mentioned above concerning the equivalence 
between the FF construction and that from the lattice 
regularization has been addressed in ref.~\cite{BNNPSW}. 
The two constructions are both non--perturbative and in both cases
it is the low energy properties that are most easily accessible.
However in the bootstrap approach it is simpler to obtain reliable
results at higher energies. For example for the Fourier
transform of the 2--point correlation function of Noether currents,
the contributions of higher $r$--particle states become significant  
at typical energies rapidly increasing with $r$. Of course,
despite the fact that in practice renormalized perturbation theoretic 
computations often seem to be quite accurate down to unexpectedly low 
energies, the property of asymptotic freedom is a statement pertaining 
to asymptotically high energies, and to establish this property in 
the bootstrap framework it is necessary to have rigorous knowledge
on the contributions of $r$--particle states with arbitrarily large $r$.
Although the latter has not yet been achieved, based on the
exact form factors up to 6 particles, Max Niedermaier and one of the 
present authors (J.B.) presented convincing evidence for remarkable
{\it scaling properties} of the higher intermediate
particle contributions~\cite{BN,BNII}, which are very powerful
because they make it possible to include 
dominant contributions from all states in the 
form factor expansion of correlation functions. Furthermore the
scaling hypothesis is completely compatible with asymptotic freedom.

Although the lattice regularization is an important non--perturbative method  
(and practically the only one available in 4d QCD), there are many 
physical phenomena
which are inaccessible in this approach for example nuclear structure functions
at small Bjorken $x$. In the framework of the bootstrap approach in 2d however,
we hope that properties of the structure functions at small
$x$ can be extracted, an insight which would hopefully also be relevant
for QCD. This hope would be realized if we could find evidence for
scaling properties of the structure functions similar to those already
observed for the 2--point function spectral densities.
 
With this goal we have initiated a project to compute the
structure functions in the O(3) sigma model, and with the
known form factors we have computed contributions 
from intermediate states up to 5 particles. To establish
their scaling properties however it is helpful to extend the list
of known form factors to higher particle number.
In this paper we discuss the calculation of the 7--particle 
form factors, which although in principle trivial, is technically 
challenging because for $r=7$ one has to deal with quite large
polynomials. We will report on the structure functions in a future
publication. Here we restrict attention to the 
contribution of the 7--particle intermediate states to 
the spin--spin correlation function,
which yields an additional test of the scaling hypothesis.
We also exhibit simple bounds on the
square of the known form factor polynomials, which if they could be
generalized to an arbitrary number of particles  
might be useful for analytic estimates to establish general properties.

\section{Some basic definitions}
The central object of attention in this paper is
the 7--particle form factor of the spin operator of the
O(3) model. This is the $r=7$ case of 
\be
\langle0\vert S^a(0)\vert a_1,\theta_1;\dots;a_r,\theta_r\rangle
=\frac{2}{\sqrt{\pi}}\,f^a_{a_1\dots a_r}(\theta_1,\dots,\theta_r),
\label{FF}
\end{equation}
where the particles are labelled by their isospin indices and
rapidities and the spin operator $S^a$ also carries an isospin
index. The normalization factor $2/\sqrt{\pi}$ is introduced here
for later convenience. The spin operator itself is normalized
by
\be
\langle0\vert S^a(0)\vert a_1,\theta_1\rangle=\delta^{aa_1}\,,
\,\,\,\,\,\,\,\,
\langle b, \chi\vert a,\theta\rangle=4\pi\delta^{ab}\delta(\chi-\theta)\,,
\end{equation}
which means that the Minkowski propagator has unit residue at
$p^2=M^2$, where $M$ is the physical mass of the O(3) particle. 

Introducing the squared form factors
\be
\delta^{ab}\, F^{(r)}(u)=\sum_{a_1\dots a_r}
f^a_{a_1\cdots a_r}(\theta_1,\dots,\theta_r)^*
f^b_{a_1\cdots a_r}(\theta_1,\dots,\theta_r)
\end{equation}
the spectral density is given by
\be
\rho^{\rm spin}(\mu) = \frac{4}{\pi}
\sum_{k=0}^\infty\rho^{(2k+1)}(\mu),
\end{equation}
where
\be
\rho^{(r)}(\mu) = \int_0^{\infty}
\frac{{\rm d} u_1 \ldots {\rm d} u_{r-1}}{(4 \pi)^{r-1}}\;F^{(r)}(u)\; 
\delta(\mu - M^{(r)}(u))\;.
\end{equation}
Here $u_j=\theta_j-\theta_{j+1}$ are rapidity differences and
$M^{(r)}(u)$ is the $r$--particle invariant mass.
The Fourier transform of the correlation function is represented as
the Stieltjes transform of the spectral density:
\be
I^{\rm spin}(p^2) = \int_0^{\infty}{\rm d}\mu\,\,
\frac{\rho^{\rm spin}(\mu)}{p^2+\mu^2}\;.
\label{Stieltjes}
\end{equation}

We will compare (the truncation of) (\ref{Stieltjes}) to the results
of perturbation theory. The 2--loop order perturbative result 
is \cite{UWolff} 
\be
p^2\, I^{\rm spin}(p^2)=\lambda_1\left\{\frac{1}{\alpha(p)}+(2+\xi_0)+
(2+\xi_0)\alpha(p)+O\left(\alpha^2(p)\right)\right\}\,.
\label{PT}
\end{equation}
Here the running coupling function $\alpha(p)$ is the solution of
\be
\frac{1}{\alpha(p)}+\ln\alpha(p)=\ln\frac{p}{M}
\end{equation}
and the parameter $\xi_0$ gives the connection between the
perturbative mass parameter $\Lambda_{\overline{\rm MS}}$ and the 
exact mass gap $M$. In the O(3) model their ratio is known exactly \cite{HN}:
\be
\xi_0=\ln \frac{M}{\Lambda_{\overline{\rm MS}}}=\ln 8-1\,\,\approx\,\,
1.07944\,.
\label{MperLambda}
\end{equation}
The overall constant $\lambda_1$ cannot be calculated in perturbation
theory, but has been determined exactly \cite{BN} using the scaling 
hypothesis (see also Sect. 5):
\be
\lambda_1=\frac{4}{3\pi^2}.
\label{lambda1}
\end{equation} 

Instead of the physical form factor (\ref{FF}) it is convenient to
consider the reduced form factors $g^a_{a_1\dots a_r}$ defined by
\be
f^a_{a_1\cdots a_r}(\theta_1,\dots,\theta_r)=\frac{\pi^{\frac{3r}{2}-1}}{2}
\Psi(\theta_1,\dots,\theta_r)\,
g^a_{a_1\cdots a_r}(\theta_1,\dots,\theta_r).
\label{redFF}
\end{equation}
Here
\be
\Psi(\theta_1,\dots,\theta_r)=\prod_{i<j}\psi(\theta_i-\theta_j)\,,
\label{Psi}
\end{equation}
with
\be
\psi(\theta)=\frac{\theta-i\pi}{\theta(2\pi i-\theta)}
\tanh^2\frac{\theta}{2}\,.
\label{psi}
\end{equation}
We note that (\ref{psi}) is the basic 2--particle form factor and
the product (\ref{Psi}) would be the complete $r$--particle
form factor if there were no isospin degrees of freedom.

The advantage of using the reduced form factors is that they are
polynomials and can (in principle) be recursively calculated
\cite{BH}. The simplest nontrivial case is $r=3$:
\be
g^a_{a_1a_2a_3}(\theta) = 
\delta^{aa_3}\delta^{a_1a_2}\,(\theta_2-\theta_1)+
\delta^{aa_2}\delta^{a_1a_3}\,(\theta_1-\theta_3-2i\pi)+
\delta^{aa_1}\delta^{a_2a_3}\,(\theta_3-\theta_2).
\end{equation}
The higher reduced form factors very rapidly become complicated
because of the large number of isospin components and rapidity
variables.

The quantity entering the spectral densities and two--point functions is   
the absolute square of the form factors, summed over the internal symmetry
indices. For the reduced form factors the corresponding quantities
are
\be
G^{(r)}(\theta_1,\dots,\theta_r)
= \frac13\sum_{a a_1\cdots a_r}
|g^a_{a_1\cdots a_r}(\theta_1,\dots,\theta_r)|^2.
\end{equation}
For the $r=3$ example we get
\be
G^{(3)}(\theta_1,\theta_2,\theta_3)
=2\left[(\theta_1-\theta_2)^2+(\theta_1-\theta_3)^2+
(\theta_2-\theta_3)^2\right]+12\pi^2.
\label{SQ3}
\end{equation}

\section{The 7--particle form factors}
The 7--particle reduced form factors 
\be
g^a_{a_1\dots a_7}(\theta_1,\dots,\theta_7)
\label{red7}
\end{equation}
are polynomials of degree 15 in the rapidity differences.
To count the independent isospin components we can obviously fix
the operator index $a=3$ since the rest of the components are related
by simple isospin transformations. Then we get 274 non--vanishing
components not counting terms related by the $1\leftrightarrow2$
isospin symmetry twice. These components can be calculated from
the known 5--particle form factors using the inhomogeneous FF axioms
\cite{BH}. The resulting polynomials contain typically
$\sim  10^4$ terms.
Actually it is not necessary to do this calculation for all
the 274 components since we can find 4 components such that the rest
can be obtained from them by using the homogeneous FF axioms
(corresponding to permutations of the rapidity variables).
After having computed all terms we checked once more that all the
FF axioms are satisfied by (\ref{red7}) and also that the coefficients
of the leading terms in the last variable (those proportional to 
$\theta_7^5$) are proportional to the 6--particle reduced form factors
as they should \cite{BN}.

After the calculation of the form factors the next step is to compute
the sum of the absolute square of the components since this is needed
in the calculation of the 7--particle contribution to the spin--spin
correlation function and other physical quantities. This calculation
is rather challenging because at intermediate steps one has to handle
quite large polynomials (of degree 30 in 7 variables
consisting of about $\sim 4\cdot 10^5$ terms).
The final result can be simplified enormously because, being
a polynomial symmetric under permutations of the rapidity variables,
it can be expanded in terms of the basic symmetric polynomials
$\sigma_1,\sigma_2,\dots,\sigma_7$ 
(defined in Eq.~(A.3) of ref.~\cite{BN}).
Written in this way, the final result consists
of only 3214 terms and is reduced to a manageable size.
\footnote{Which is however still too large to usefully reproduce here.}    

To be sure that we have obtained the correct result for the
square $G^{(7)}(\theta_1,\dots,\theta_7)$ we have performed
a number of checks. First we checked that the leading terms in the
last variable (the coefficient of $\theta_7^{10}$) are twice the 
analogous 6--particle square $G^{(6)}(\theta_1,\dots,\theta_6)$ \cite{BN}.
Next we checked that the overall leading terms (the terms of
total degree 30) agree with those of a simple symmetric polynomial
$P^{(7)}(t)$ \cite{BN}. This is defined by (for $r$ particles)
\be
P^{(r)}(t)=\sum_{{\rm perms}} \, P^{(r)}_0(t),
\label{perm}
\end{equation}
where
\be
P^{(r)}_0(t)=\prod_{j-i>1}\,\left[(\theta_i-\theta_j)^2+t\pi^2\right].
\label{t}
\end{equation}
In (\ref{perm}) the sum extends over the $r!$ permutations of the 
rapidity variables $\theta_1,\dots,\theta_r$ and in (\ref{t}) the
correction terms depending on the parameter $t$ do not affect
the overall leading terms, they are included here for later
convenience. Finally, we verified that $G^{(7)}(\theta_1,\dots,\theta_7)$
vanishes at the points \cite{BN}
\be
\theta_1-\theta_2=\theta_2-\theta_3=i\pi,\qquad\qquad
\theta_j={\rm arbitrary}\qquad j>3.
\end{equation}

The fact that $G^{(7)}$ passed all these nontrivial tests gives us
some confidence in the correctness of our results.

\section{The spin--spin correlation function}
Having calculated the square of the 7--particle form factors
the corresponding contributions to the spin--spin correlation function
and the spectral density can now be computed straightforwardly.
The only difficulty is that to get correct numerical results
one has to use high precision arithmetic in order to avoid
huge rounding errors in calculating $G^{(7)}$.
As explained above, to put it into a manageable form, we expressed it
in terms of the basic symmetric polynomials 
$\sigma_1,\dots,\sigma_7$.
Although it is a sum of absolute squares and thus 
it is obviously positive, it is not manifestly positive in this form.
Actually we experienced large cancellations between positive and
negative contributions. After rescaling all rapidities by $\pi$ all
coefficients of the polynomial become integers and we can illustrate
the large degree of cancellation by considering the following
integer rapidities:
\be
\theta_1=12,\quad
\theta_2=11,\quad
\theta_3=10,\quad
\theta_4=9,\quad
\theta_5=8,\quad
\theta_6=7,\quad
\theta_7=0.
\end{equation}
We can exactly calculate the sum of positive and negative
contributions in this case and we get
\bea
{\rm positive\ part}&:&\quad 265129842869261203568
67283491794696305
689824\nonumber\\
{\rm negative\ part}&:&\quad 265129842869261203568
55853186993819121
689824\nonumber\\
{\rm total}&:&\quad \phantom{265129842869261203568}
11430304800877184
000000\nonumber
\end{eqnarray}
showing a 21--digit cancellation here!

\begin{figure}[htb]
\begin{flushleft}
\leavevmode
\epsfxsize=140mm
\epsfbox{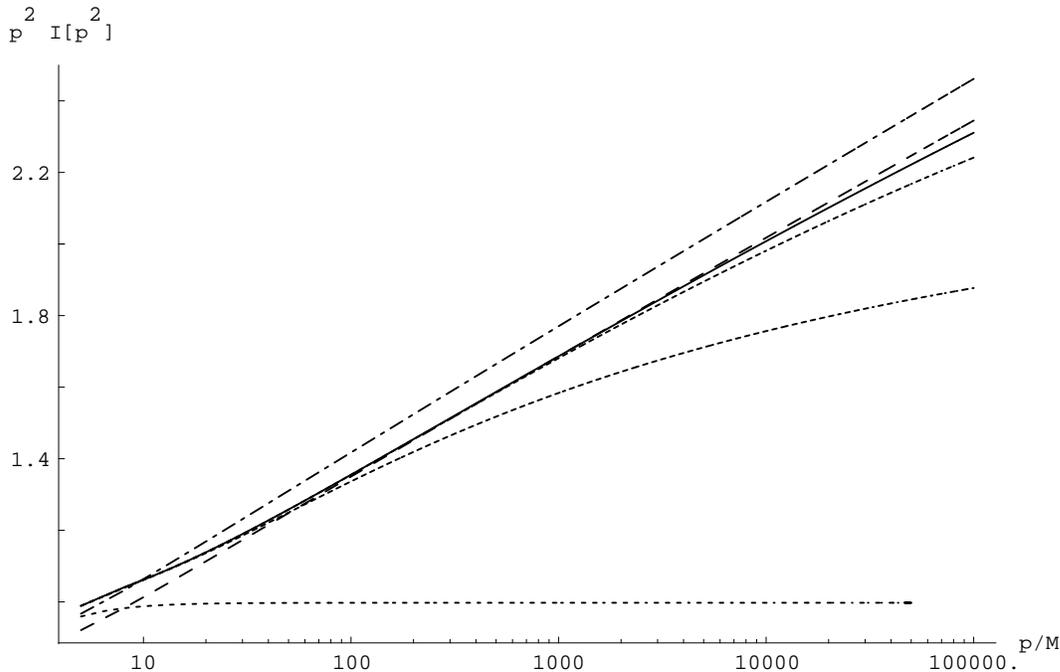}
\end{flushleft}
\caption{{\footnotesize
Plot of $p^2 I^{\rm spin}(p^2)$ against $\ln(p/M)$, showing a
comparison of the form factor approach to 2--loop perturbation theory.
The solid curve is the FF approximation to $p^2 I^{\rm spin}(p^2)$ obtained by
including all intermediate states with $\le 7$ particles. The lower curves
show the analogous approximations with $\le5,\le3,1$ particles 
respectively. The normalization of the (dashed) PT curve is fixed 
according to (\ref{lambda1}).
Finally the top curve is the PT curve multiplied by a factor $1.05$.
}} 
\label{SPIN7bw}
\end{figure}

We have computed the 7--particle contribution to the spin--spin
correlation function using the VEGAS integration routine,
and a subroutine computing $G^{(7)}$ invoking quartic precision (32 digit)
arithmetic. 
The results, compared with the prediction of two--loop perturbation theory,
are shown in Figure~\ref{SPIN7bw}. Note that here we have no free 
parameters
at our disposal in the perturbative calculation as would be the case
in most other models. As discussed above, we know the exact relation between
the perturbative $\Lambda$ parameter and the particle mass $M$ 
and also the absolute normalization of the perturbative curve is
available in this model. The form factor results are in very good
agreement with perturbation theory in the expected (high) energy range.
The small deviation for energies $p/M\sim 10^4$ 
can be accounted for by the contribution of $r>7$ intermediate particles.
To illustrate the fact that this good agreement is quite nontrivial,
in Figure~\ref{SPIN7bw} we also show the perturbative result 
with the overall factor changed arbitrarily by 5\%.

\section{Scaling}
The 7--particle results also corroborate the scaling hypothesis \cite{BN}
for the spectral densities. To study this aspect we introduce the
modified $r$--particle spectral density depending on the logarithmic 
variable $x$ by the definition
\be
\mu\,\rho^{(r)}(\mu)=R^{(r)}(x)\,,\qquad\qquad \mu=Me^x.
\end{equation}
These are defined for all $r\ge2$, where the cases with $r$ even refer
to the spectral densities of the two--point function of the
isospin Noether current. 
The graph of this function is a bell--shaped curve starting as zero
at $x=\ln r$, reaching its maximum $M^{(r)}$ at $x=\xi^{(r)}$ and then slowly
decreasing for larger $x$. Let us introduce the rescaled spectral
density $Y^{(r)}$ by  
\be
Y^{(r)}(z)=\frac{1}{M^{(r)}}\,R^{(r)}\left(\xi^{(r)}z\right).
\end{equation}

\begin{figure}[bh]
\begin{flushleft}
\leavevmode
\epsfxsize=150mm
\epsfysize=90mm
\epsfbox{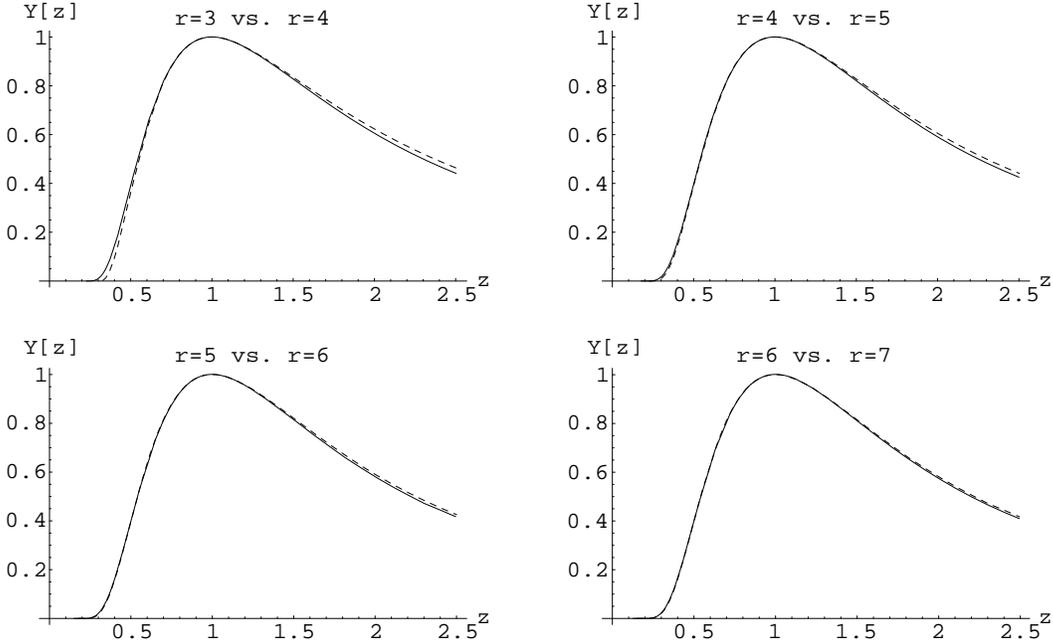}
\end{flushleft}
\caption{{\footnotesize
Illustration of the self-similarity property of the rescaled 
spectral densities. The plots show $Y^{(r)}(z)$ (dashed) 
compared with $Y^{(r+1)}(z)$ (solid) for $r=3,4,5,6$.}}
\label{SSIMCURR7}
\end{figure}

It has been found (based on the study of up to 6 particles) that 
the shape of the rescaled spectral density and the parameters
$\xi^{(r)}, M^{(r)}$ satisfy self--similarity 
\be
\lim_{r\to\infty} Y^{(r)}(z)\to Y(z)
\end{equation}
with universal shape function $Y(z)$ and asymptotic scaling,
\be
M^{(r)}\sim M_*\,r^{-\gamma}\,,\qquad\qquad
\xi^{(r)}\sim \xi_*\,r^{1+\alpha}
\end{equation}
for large $r$, with some coefficients $M_*,\xi_*$ and
exponents $\gamma, \alpha$. Whereas the properties of the form factors
are consistent with $\gamma=1$ only, the other exponent can only be 
determined numerically with result $\alpha=0.27$.
For those readers unfamiliar with this model we  
point out the amusing fact that
the self--similarity holds for all $r\ge2$ and thus interrelates
the spectral functions of two different isovector operators,
the spin field and the conserved vector current.
Similarly (25) is equally valid for even and odd $r$ values,
provided we use the normalization introduced in (1).

Self--similarity continues to be satisfied very well as demonstrated
in Figure \ref{SSIMCURR7}. To test asymptotic scaling, we
used the fitted numerical values based on results for up to
6 particles to \lq\lq predict" for the $r=7$ case
\be
M^{(7)}_{{\rm sc}}=0.03188\qquad\qquad{\rm and}\qquad\qquad
\xi^{(7)}_{{\rm sc}}=17.73\,,
\end{equation}
and compared it to the values directly determined from our 7--particle
results
\be
M^{(7)}=0.03189\qquad\qquad\,{\rm and}\qquad\qquad
\xi^{(7)}=17.77.
\end{equation}

Similarly for the integrals
\be
c^{(r)}=\int_M^\infty\,{\rm d}\mu\,\rho^{(r)}(\mu)
=\int_0^\infty\,{\rm d}x\,R^{(r)}(x)\qquad\qquad
h^{(r)}=\int_0^\infty\,\frac{{\rm d}x}{x}\,R^{(r)}(x)
\label{integrals}
\end{equation}
we predict
\be
c^{(7)}_{{\rm sc}}=1.464\qquad\qquad{\rm and}\qquad\qquad
h^{(7)}_{{\rm sc}}=0.04879\,,
\end{equation}
and actually get
\be
c^{(7)}=1.46(1)\qquad\qquad\,{\rm and}\qquad\qquad
h^{(7)}=0.0488(1).
\end{equation}

\section{Ultra--positivity}
Because of the large cancellation between positive and negative terms
in the representation of $G^{(r)}$ in terms of symmetric polynomials,
it is natural to ask if there exists an alternative representation
that is manifestly positive. It is indeed possible to find such
representations. If we arrange the rapidities in decreasing order
(which is always possible since the polynomial is symmetric under
permutations) then all $u_j=\theta_j-\theta_{j+1}$ rapidity
differences are positive and we found (for all available
particle numbers $3\leq r\leq7$) that if we expand $G^{(r)}$
in terms of these differences then all coefficients are positive.
Moreover, it is possible to find upper and lower bounds both of which
are of the simple form (\ref{perm}) such that, expanded in terms of
the rapidity differences $u_j$, each and every term in the expansion
of $G^{(r)}$ has smaller coefficient than the corresponding one of
the upper bound $P^{(r)}(t^{(r)}_u)$ and similarly larger than the
coefficient of the corresponding term of the lower bound  
$P^{(r)}(t^{(r)}_l)$.

The value of the parameter characterizing the lower bound is
uniformly $t^{(r)}_l=2$ for all cases  $3\leq r\leq7$, whereas
the $t^{(r)}_u$ giving the lowest upper bound are
\be
t^{(3)}_u=2,\quad
t^{(4)}_u=9/4,\quad
t^{(5)}_u=2.434,\quad
t^{(6)}_u=2.576,\quad
t^{(7)}_u=2.688.
\end{equation}
The last three values are numerically approximate 
\footnote{e.g. the exact value for the case $r=5$ is $t^{(5)}_u=208^{1/6}$}
and come from the requirement
that the constant term of $P^{(r)}(t^{(r)}_u)$ be larger than
that of $G^{(r)}$. This is clearly necessary, but at least in the
available cases for $r\ge5$ it is also sufficient
\footnote{The case $r=4$ is an exception since
the requirement in this case yields a value $(34/3)^{1/3}=2.246$,
which is not quite sufficient.}.

The integrals over the spin and current spectral densities are 
given as sums of the integrals defined in (\ref{integrals}):
\begin{eqnarray}
C^{{\rm spin}}&=&\int_M^\infty\,{\rm d}\mu\,\rho^{{\rm spin}}(\mu)
=\frac{4}{\pi}\sum_{k=0}^{\infty}c^{(2k+1)}\,,
\nonumber
\\
C^{{\rm curr}}&=&\int_M^\infty\,{\rm d}\mu\,\rho^{{\rm curr}}(\mu)
=\sum_{k=0}^{\infty}c^{(2k+2)}\,.
\label{infty} 
\end{eqnarray}
An outstanding question is whether these are finite or infinite in 
the FF construction. Certainly renormalized perturbation 
theory predicts that $C^{{\rm spin}}=\infty=C^{{\rm curr}}$.
Also the validity of the scaling hypothesis requires this because 
in this scenario $c^{(r)}$ grows as $\sim r^\alpha$; the numerical 
evidence thereof is shown in Figure~\ref{cr}.

If the bounds we found above for $3\leq r\leq7$ 
continue to be true for all $r$,
then the simple structure of (\ref{perm}) allows to study the structure
of the correlation function analytically. The existence of an upper
limit of simple form may help proving the existence of the correlation
function whereas the existence of the lower bound may facilitate the
construction of a proof that $C^{{\rm spin}}=\infty=C^{{\rm curr}}$
independent of the validity of the scaling hypothesis.

\begin{table}
\begin{center}
\begin{tabular}[t]{c||c|c|c|c|c|c|c}
$r$ & 1 & 2 & 3 & 4 & 5 & 6 & 7 \\
\hline
$c^{(r)}$    & 0.785 & 1.009 & 1.140 & 1.242 & 1.327 & 1.400 & 1.46 \\
$c^{(r)}(2)$ &       &       & 1.140 & 1.206 & 1.229 & 1.225 & 1.200 \\
\hline
\\
$r$ & 8 & 9 & 10 & 11 &  &  &  \\
\hline
$c^{(r)}(2)$ & 1.164 & 1.118 & 1.067 & 1.013 & & &\\
\end{tabular}
\end{center}
\caption{{\footnotesize Numerical values of the integrals 
$c^{(r)}, c^{(r)}(2)$. Numerical errors are estimated as $\pm1$ on the 
last digit quoted.}}
\end{table}

We are not going to discuss these questions further in this paper,
but to illustrate the usefulness of the existence of the simple lower bound
we have calculated the integrals $c^{(r)}(2)$. These are defined
analogously to the ones in (\ref{integrals}) using $P^{(r)}(2)$
instead of $G^{(r)}$. The simplicity of the integrand allows us to
calculate these integrals numerically quite effortlessly 
\footnote{This is because one can represent $c^{(r)}(2)$
as an integral with the integrand just involving the polynomial $P_0^{(r)}(2)$
in Eq.~\ref{t} (instead all its permutations), at the cost of extending
the limits of integration over the $u_i$ from $-\infty$ to $+\infty$.}
up to $r=11$. The results of the numerical integration are given in Table~1.

Although the integrals $c^{(r)}(2)$ are (apparently) decreasing with
$r$, it is possible that they are decreasing slow enough to make
the series (\ref{infty}) diverge. Indeed, as seen in Figure~\ref{cr},
$rc^{(r)}(2)$ is monotonically increasing for all $r$ evaluated 
up to now.
\vspace{1cm}

\begin{figure}[htb]
\begin{flushleft}
\vskip 10mm
\epsfxsize=140mm
\epsfbox{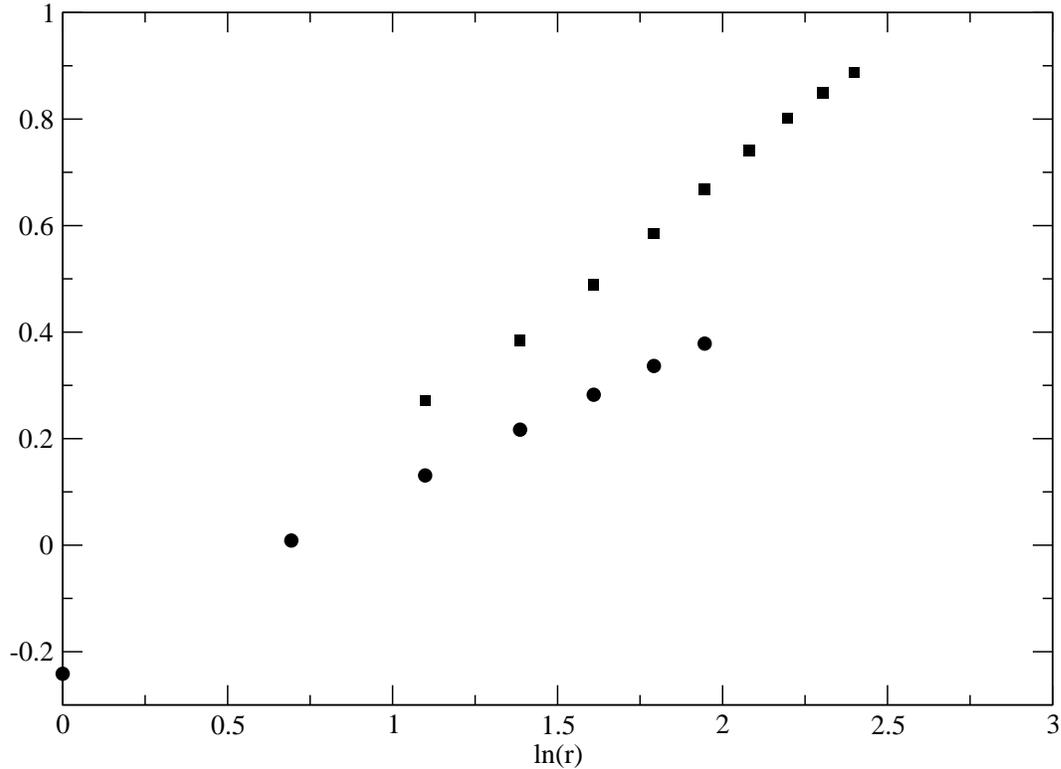}
\end{flushleft}
\caption{{\footnotesize
Values of $\ln c^{(r)}$ (circles) and of the product 
$rc^{(r)}(2)/(4\pi)$ (squares). The error on $\ln c^{(7)}$ is 
approximately the radius of the circle. 
}}
\label{cr} 
\end{figure}


\vspace{1cm}
{\tt Acknowledgements}

\noindent 
J. B. wishes to thank the Max--Planck--Institut f\"ur Physik in Munich
where most of this work has been done, for hospitality.
This investigation was supported in part by the 
Hungarian National Science Fund OTKA (under T030099 and T034299).

\vspace{3cm}


\end{document}